# Identifying potentially induced seismicity and assessing statistical significance in Oklahoma and California

[1,2]Mark McClure, [3]Riley Gibson, [1]Kit-Kwan Chiu, and [4]Rajesh Ranganath

[1]Department of Petroleum and Geosystems Engineering, The University of Texas at Austin (former)

[2]McClure Geomechanics LLC

[3]Department of Computer Science, The University of Texas at Austin (former)

[4]Department of Computer Science, Princeton University

mark.w.mccl@gmail.com; rajeshr@cs.princeton.edu

**Abstract.** We develop a statistical method for identifying induced seismicity from large datasets and apply the method to decades of wastewater disposal and seismicity data in California and Oklahoma. The method is robust against a variety of potential pitfalls. The study regions are divided into gridblocks. We use a longitudinal study design, seeking associations between seismicity and wastewater injection along time-series within each gridblock. The longitudinal design helps control for non-random application of wastewater injection. We define a statistical model that is flexible enough to describe the seismicity observations, which have temporal correlation and high kurtosis. In each gridblock, we find the maximum likelihood estimate for a model parameter that relates induced seismicity hazard to total volume of wastewater injected each year. To assess significance, we compute likelihood ratio test statistics in each gridblock and each state, California and Oklahoma. Resampling is used to empirically derive reference distributions used to estimate p-values from the likelihood ratio statistics. In Oklahoma, the analysis finds with extremely high confidence that seismicity associated with wastewater disposal (or other related activities, such as reservoir depletion) has occurred. In California, the analysis finds that seismicity associated with wastewater disposal has probably occurred, but the result is not strong enough to be conclusive. We identify areas where temporal association between wastewater disposal and seismicity is apparent. Our method could be applied to other datasets, extended to identify risk factors that increase induced seismic hazard, or modified to test alternative statistical models for natural and induced seismicity.

## 1. Introduction.

**1.1. Background.** In recent years, certain regions across the United States have experienced a large increase in seismicity relative to the historical baseline. The increase in seismicity has been linked to underground wastewater disposal wells, which are primarily associated with oil and gas production (Frohlich, 2012; Hitzman, 2013; Keranen et al., 2014; Weingarten et al., 2015; Walsh and Zoback, 2015).

Hints to induced seismicity are found in temporal and spatial correlation between seismicity and human activity with the potential to induce seismicity (Davis and Frohlich, 1993; Frohlich et al., 2016). However, because of the large number of natural earthquakes and the large number of wastewater disposal wells, it is not always clear whether an instance of apparently induced seismicity is coincidental. In many cases, the temporal and spatial link between injection and seismicity is so strong that induced seismicity is unambiguous (Raleigh et al., 1976; Häring et al., 2008; Frohlich, 2012; Kim, 2013). But many factors can create ambiguity in the identification of induced seismicity: high background seismicity rate, uncertain event locations (especially in areas with poor seismic coverage), time-lag between the beginning of injection/extraction and the onset of seismicity, and spatial separation between injection/extraction and the onset of seismicity (Davis and Frohlich, 1993; Frohlich et al., 2016).

---

[*] Revisions are: (1) significant modification of the discussion in Section 1.3, (2) addition of one new paragraph in Section 1.1, (3) correction of a typo in the list of authors, and (4) slight rewording of the first sentence in the abstract.

To address these issues, we develop a statistical approach to identifying induced seismicity in large datasets while also assessing statistical confidence at the local and regional scale. Our method is designed to avoid potential pitfalls associated with statistically identifying induced seismicity (discussed in Section 1.3). We apply our algorithm to study induced seismicity associated with wastewater disposal in two states: Oklahoma (2000-2013) and California (1980-2013). By aggregating large, comprehensive datasets including earthquakes and injection wells over many years, we are able to assess whether associations between injection and seismicity occur more often than would be expected by chance. We identify specific areas in both states where association between injection and seismicity is apparent. In future work, our method could be extended to identify factors that increase the risk of induced seismicity.

Assessing statistical significance has practical and policy applications when decisions are made about whether to curtail injection in response to potentially induced seismicity. For example, Hornbach et al. (2015) found that seismicity was induced by wastewater injection near Azle, TX. However, the Texas Railroad Commission declined to order the operators to cease injection, finding that "the evidence in the record is insufficient to conclude that injection is likely to be or determined to be contributing to seismic activity" (Dubois and Enquist, 2015). A statistical assessment of the probability of coincidental correlation could have helped make the decision about whether to shut-in the well.

The data in this study is restricted to focus solely on wastewater injection wells, not injection wells classified for improved oil recovery or geothermal energy extraction. Injection for improved oil recovery and geothermal energy extraction should be treated as separate categories. Induced seismicity is relatively common in the exploitation of geothermal energy (Evans et al., 2012; Trugman et al., 2016). Induced seismicity is less likely during injection for improved oil recovery because injection and production occur simultaneously in the same formation, though at least one potential instance has been documented in the literature (Gan and Frohlich, 2013).

Hydraulic fracturing for oil and gas recovery has only rarely been suspected of causing induced seismicity. The volume of fluid injected during hydraulic fracturing is much less than in long-term injection wells, and so the induced seismicity hazard is lower. A notable exception is in the Western Canada Sedimentary Basin, where Atkinson et al. (2016) found an association between seismicity and hydraulic fracturing (though there is some ambiguity in distinguishing between the effect of hydraulic fracturing and nearby disposal wells). Hydraulic fracturing is a qualitatively different process from long-term water injection, and so should be treated as a separate category. Hydraulic fracturing is not included in our study, but in future work, our approach could be used to test for associations between hydraulic fracturing and seismicity.

**1.2. Mechanisms of induced seismicity.** Induced seismicity occurs when human activity changes the pressure or stress on a fault, causing it to slip. The Coulomb failure criterion characterizes fault slip (Jaeger et al., 2007):

$$|\tau| \leq \mu_f(\sigma_n - P), \tag{1}$$

where $\tau$ is the shear stress, $\mu_f$ is the coefficient of friction, $\sigma_n$ is the normal stress, and $P$ is the fluid pressure. A fault will slip once the magnitude of its shear stress exceeds the frictional resistance to slip (given by the right-hand-side of Equation 1). Rate and state friction models describe variation of the coefficient of friction over time and explain time-delayed nucleation, aseismic (slow) slip, and other processes (Segall, 2010; McClure and Horne, 2011).

Processes that increase $P$ or $|\tau|$ or decrease $\mu_f$ or $\sigma_n$ can bring a fault closer to slip, potentially inducing seismicity. Injection causes pressure increase, which weakens friction, resulting in slip. Many human activities can induce seismicity, including fluid injection, fluid extraction, surface reservoir impoundment, and mine excavation (McGarr et al., 2002).

**1.3. Challenges for identification of induced seismicity.** In this section, we list a variety of challenges for statistically identifying induced seismicity.

The first challenge is that injection wells are not located at random, and background seismicity rate is spatially variable. In datasets with non-random assignment, confounding factors, those that influence both injection and seismicity, have the potential to create the appearance of a causal relationship (Chapter 6 from Pearl, 2009). As a hypothetical example, hydrocarbon deposits tend to occur in sedimentary basins, which are extensional regions of the crust with relatively relaxed stress. Therefore, it might be hypothesized that hydrocarbon deposits (and associated water injection) will tend to be found in regions of low natural seismicity. If so, this confounding factor might result in the observation that seismicity is lower on average in the vicinity of water disposal wells than in areas without water disposal wells.

The second challenge is spatial clustering of injection wells and earthquakes. Spatial clustering creates ambiguity about causality and complicates the assessment of statistical significance. For example, Weingarten et al. (2015) performed an analysis of seismicity and injection data across the central United States. Wells were considered "associated" with seismicity if an earthquake occurred within 15 km during the period that the well was actively injecting. Weingarten et al. (2015) estimated confidence intervals using bootstrap resampling, an approach that is only valid if the observations at each well are independent. However, multiple wells may be within 15 km of the same earthquake, which creates correlation between the observations. The closer together injection wells are located, the greater the correlation between their observations. The consequence of neglecting correlation between observations is overestimation of statistical significance.

The third challenge is the risk of false positives. Testing large number of hypotheses increase the probability of finding a coincidental associations (Benjamini and Hochberg, 1995). For example, Brodsky and Lajoie (2013) sought cross-correlations across a wide range of one-month time lags between a declustered earthquake catalog and total production volume, injection volume, and net production volume at the Salton Sea geothermal field. The analysis was performed on three separate time-periods and also over the entire duration of the dataset. Overall, twelve different analyses were performed (four time periods multiplied by three types of volume relationships), and in each analysis, time lags from zero up to at least 170 months were tested, for a total of at least 170×12 = 2040 different hypotheses tested. In some parts of the data, observed relationships appear meaningful. For example, from 1982 to 1991, cross-correlation values greater than 0.9 were observed at a time-lag of 0 months for all three types of volumes tested: production, injection, and net production volume. But during the other time-periods, reported maximum cross-correlations were typically less than 0.25 and occurred at large, unphysical time-lags, such as 89 or 170 months. These cross-correlation values were almost certainly coincidental, artifacts emerging from the large number of hypotheses tested.

The fourth challenge is that the detection sensitivity of the seismic array is spatially variable and changes as the seismic array evolves over time (Schorlemmer and Woessner, 2008). Because changes to the seismic array are nonrandom, this has the potential to be a confounding factor that creates a spurious relationship in the data. For example, Goebel et al. (2015) performed a statistical analysis to seek evidence of wastewater-induced seismicity in the vicinity of 1400 wells in central California. The completeness magnitude is defined as the magnitude above which nearly all earthquakes will be detected. Goebel et al. (2015) noted that the completeness magnitude was 2.0 but used the entire seismic catalog (including events below the magnitude of completeness) and also included events from a separate (more sensitive) waveform-relocated catalog that was available in part of the study area. The nonuniform earthquake detection sensitivity could create a spurious correlation between injection and seismicity if: (a) injection wells tend to be located in the section of the study area where the (more sensitive) waveform-relocated catalog was used, (b) seismic sensors tend to be placed in regions of greater human activity, or (c) injection activity has increased over time and the sensitivity of the seismic catalog has increased over time. Another problem is that spatially and temporally variable sensitivity of the earthquake catalog introduces correlation between observations (because earthquakes are more likely to be detected in areas with greater detection sensitivity), which complicates assessment of significance.

The fifth challenge is the difficulty in avoiding implicit assumptions when assessing significance. For example, Goebel et al. (2015) performed several statistical tests (collectively called the OISC method) to quantify association with seismicity on a per-well basis. To control for the possibility of false positives, they performed an assessment on the overall dataset and calculated a p-value of 0.04. However, rather than performing the assessment on the results from the full OISC method, Goebel et al. (2015) performed a separate, simplified analysis on only the largest events in the earthquake catalog. Tests of statistical significance should be performed using the same analysis and dataset as the reported results.

The simplified analysis was performed by bootstrapping from 1,000 synthetic earthquake catalogs generated by assuming spatially uniform seismic hazard. This is problematic because actual earthquake hazard is spatially variable. Relatively large earthquakes in a region tend to cluster around the largest faults. Well locations are also spatially correlated. Spatial clustering of wells and earthquakes increases the variance of the null distribution and needs to be taken into account when assessing significance. In contrast, Atkinson et al. (2016), who performed a study of induced seismicity due to hydraulic fracturing in the Western Canada Sedimentary Basin, assessed significance by generating synthetic earthquake data with a technique that accounts for spatial and temporal variability in natural seismicity.

Our method (described in Section 2) is designed to overcome these difficulties. We perform our analysis on non-overlapping spatial blocks so that each earthquake is uniquely mapped to a single unit in the analysis (in contrast to approaches that seek association between earthquakes and individual wells). We test for temporal associations between injection and seismicity within geographic areas, rather than seeking cross-sectional relationships between geographic areas (the latter approach can be confounded by nonrandom assignment). We assess significance using resampling in a way that preserves underlying statistical properties of the seismicity and injection data. We use only earthquakes greater than the magnitude of completeness. To account for the number of hypotheses tested, we aggregate the results in each block into an overall assessment of statistical significance within each state.

## 2. Methods.

**2.1. Statistical model and data.** We divide California and Oklahoma into gridblocks of 0.2° latitude and longitude on each side (roughly 14×11 mi$^2$). We use a longitudinal study design, comparing injection and seismicity along time within each gridblock, rather than comparing between gridblocks, as in a cross-sectional study. The longitudinal study design mitigates confounding factors that might cause wastewater injection to be high where natural seismicity is high or low.

The ANSS Comprehensive Catalog is used for California and filtered to include only events greater than or equal to magnitude 3.0, approximately the completeness magnitude during the study period (Schorlemmer and Woessner, 2008), and events with depth of 7.0 km or less. Well injection volumes include only wells classified as "WD" (for water disposal) and are taken from the Division of Oil, Gas & Geothermal Resources of the California Department of Conservation.

The seismic catalog for Oklahoma is obtained from The Oklahoma Geological Survey's Seismic Monitoring Program and screened to include events above 2.9, the magnitude of completeness during the study period. The events were not screened for depth because of concern over the accuracy of the depth estimates. The injection data is taken from the database assembled by Weingarten et al. (2015), which was assembled from data provided by the Oklahoma Corporation Commission Oil and Gas Division.

A limitation of our approach is that injection in one gridblock could cause seismicity in an adjacent gridblock (or location error could cause the event to be located in an adjacent gridblock), which the method would not detect. The gridblocks are fairly large, which increases the probability that injection and potentially induced seismicity will occur in the same gridblock. If our method was applied with smaller gridblocks, it would be more important to account for discretization effects.

Several strategies could be considered in future work to address discretization effects, such as partial assignment of events and/or injection volumes between neighboring gridblocks, a modification of the model (Equation 2) to explicitly account for injection volumes in neighboring blocks, or inclusion of spatial correlation in the Normal distributions used in the model (Equation 2). To test sensitivity to the grid, we performed the analysis twice in each state, using sets of gridblocks offset by 0.1° (Section 3.5). The results were similar in both cases.

It is assumed that background seismicity rate and induced seismicity hazard are spatially variable but constant over time. Induced seismicity hazard might change over time due to changes in operational practices or other factors, and so this is a simplifying assumption.

We define a flexible and nonlinear distribution to describe natural and induced seismicity. We first set up some notation. Let $Poi(z)$ denote a draw from the Poisson distribution, with mean $z$, and let $N(0, \sigma)$ be a draw from the Normal distribution with mean of zero and standard deviation of $\sigma$. Next we define our observations. Let $y_{ij}$ be the number of earthquakes occurring in block $i$ in year $j$, and let $x_{ij}$ be the cumulative volume of fluid injected. The parameters of the distribution are $\sigma_i$, expressing the variability of seismicity from year to year, $\mu_i$, related to the rate of natural seismicity, $\beta_i$, expressing the relationship between injection volume and induced seismicity, $a$, expressing the degree of temporal correlation for $y$, and $\sigma_{II}$, expressing the variability of temporal correlation. Given this, the observations follow

$$y_{ij} = Poi(e^{N(0,\sigma_i)}(\mu_i + \beta_i x_{ij}) + a y_{i,j-1} e^{N(0,\sigma_{II})}). \qquad (2)$$

The Poisson distribution is additive: the sum of two Poisson is a Poisson with the sum of rates. Therefore, Equation 2 implies three sources of earthquakes: the background rate, the temporal clustering rate, and the water injection induced rate, which may be zero if $\beta$ is zero. The exponentiated realizations from Normal distributions inside the Poisson mean overdisperse these components to match the dispersion present in observed seismicity.

We choose not to apply a temporal declustering algorithm to the earthquake catalog. For example, the Epidemic Type Aftershock Sequence (ETAS) model is commonly used for inferring background seismicity rates (Brodsky and LaJoie, 2013; Trugman et al., 2016). However, different declustering algorithms have relative advantages and disadvantages, and there are certain phenomena, such as earthquake swarms, that are not well-described by the declustering paradigm of mainshock-aftershock sequences (van Stiphout et al., 2012). Applying Equation 2 directly on the raw data minimizes embedded assumptions and avoids reduction of statistical power due to removal of data. The overdispersion created by the exponentiated Normal distributions and the temporal correlation term in Equation 2 account for the temporal clustering created by aftershocks and earthquake swarms. The overall approach applied in this study could be used with any statistical model of earthquake occurrence for each gridblock, including the ETAS model, by replacing Equation 2 with an alternative.

Wastewater disposal is the subsurface activity associated with hydrocarbon production that is most likely to induce seismicity (Hitzman, 2012). However, reservoir depletion, hydraulic fracturing, and other activities are also capable of causing induced seismicity (McGarr et al., 2002). These other processes are not considered directly in our statistical model, but water disposal wells will tend to be in the vicinity of producing hydrocarbon reservoirs. Therefore, the $\beta$ parameter should be interpreted as representing the overall association between water disposal and other related activities, not only the effect of water injection.

Equation 2 is a purely statistical relation and is not based on an attempt to mimic particular physical processes. Equation 2 satisfies several qualitative requirements needed to reasonably describe the data. The model must yield nonnegative integer values, because it predicts the number of earthquakes in a block in each year, and it must be flexible enough to handle overdispersed seismicity observations. Because the seismic catalog is not declustered, the number of earthquakes from year to year is highly variable. The overdispersion in the model allows it to reproduce the high variability in the data.

The model must be able to capture temporal correlation in $y$. Temporal correlation can occur, for example, if a large earthquake occurs near the end of the year, and many of its aftershocks occur in the subsequent year. But there is large variability in this temporal correlation. If a large earthquake occurs at the beginning of a year, its aftershock sequence may be finished by the end of the year, and in this case, there may not be an unusually high number of earthquakes the following year. After a year with $y_{i,j-1}$ earthquakes (with $y_{i,j-1}$ elevated significantly above baseline in that block), values of $y_{ij}$ observed in the data range from near zero to values greater than or equal to $y_{i,j-1}$. Equation 2 supports overdispersion via the exponentiated draw from the Normal inside the Poisson distribution. Therefore, the temporal correlation term must also be overdispersed - capable of capturing correlation, but also capturing the large variability in correlation that is actually observed. Otherwise, outlier years showing very low correlation could have a disproportionately large effect on the results.

The temporal correlation is captured by the $y_{i,j-1}$ term in Equation 2 with parameters $a$ and $\sigma_{II}$ controlling the temporal correlation and temporal dispersion respectively. These parameters are assumed to be the same in all gridblocks and in both study areas - California and Oklahoma. The values of 0.047 and 1.33 are used for $a$ and $\sigma_{II}$, respectively. As described in Section 2.4, these values are the maximum likelihood estimates calculated on observations from the 531 gridblocks in California that had seismicity but no wastewater disposal.

**2.2. Likelihood ratio test and resampling to calculate p-values.** Two models are constructed: a null model with likelihood $L_0$, which assumes there is no relationship between injection and seismicity ($\beta_i = 0$), and an alternative model with likelihood $L_1$, which assumes $\beta_i > 0$.

For each block, the maximum likelihood estimates for $\beta_i$ and $\mu_i$ are calculated for the $L_1$ model, and the maximum likelihood estimate for $\mu_i$ is calculated for the $L_0$ model. The calculation of the model likelihoods is discussed in Section 2.3. We compare the likelihoods of the two models with the likelihood ratio test to assess statistical significance.

The likelihood ratio test statistic $D$ is:

$$D = 2\ln\left(\frac{L_1}{L_0}\right). \tag{3}$$

With a sufficiently large dataset and without any constraints on $\beta_i$, the test statistic $D$ would be distributed according to the chi-squared distribution with degrees of freedom equal to the difference in the number of parameters in the $L_1$ model and the $L_0$ model (Wilks, 1938). However, in our study, the number of datapoints in each block is relatively small: the number of years of observations is 14 in Oklahoma and 34 in California. Because the number of datapoints is relatively small and $\beta_i$ is constrained, we use resampling to empirically generate a reference distribution for estimating p-values, rather than using the chi-squared distribution.

To estimate p-values using resampling, we generate synthetic datasets in each block, using a procedure to ensure that the null hypothesis is true. The datasets are created using the following procedure: (1) within each state, injection data is randomly permuted between blocks (using only blocks that had both injection and seismicity); (2) the injection data is offset by a random number of years; and (3) the seismicity observations are left unchanged. The injection data temporal offset is cyclical, so that offset past the last year of the data is reset back to the start of the data. This procedure is repeated 60-90 times to create a population of datasets (for which the null hypothesis is true) that can be used to generate a reference distribution.

The resampled data preserves properties in the injection data such as the marginal distributional properties because the injection time-series come from other blocks in the state. The resampling technique of shifting the years preserves any temporal correlation in the real data.

For each resampled dataset, we compute maximum likelihood estimates for both the $L_1$ and $L_0$ models and calculate the test statistic $D$. The p-value is then calculated as $(n+1)/m$, where $n$ is the number of resamples with $D$ greater than the value of $D$ from the data, and $m$ is the total number of resamples. In a small number of blocks (nearly all in Oklahoma), the value of $D$ from the data is lower than in any of the resampled datasets. In this case, the estimated p-value is an upper bound, and the true p-value is lower.

In a standard likelihood ratio test, the $L_1$ likelihood is maximized with unconstrained optimization of the model parameter(s). In contrast, we impose the condition that $\beta_i$ is nonnegative (equivalent to assuming that injection cannot inhibit seismicity), which complicates the calculation of p-values. As discussed by Molenberghs and Verbeke (2007), if a univariate likelihood ratio test is performed with the constraint that the model parameter must be nonnegative, then if the null hypothesis is true, the resulting values of $D$ will be distributed according to a mixture distribution with a 50% chance of drawing $D$ equal to zero (corresponding to cases where an unconstrained analysis would yield negative $\beta_i$). Values where $D$ equals zero in the constrained optimization correspond to cases where $\beta_i$ would be nonpositive in an unconstrained optimization. Because of the constraint, if $D$ is equal to zero ($\beta_i$ equal to zero in the constrained optimization), this indicates a p-value between 0.5 and 1.0, and $D$ cannot be mapped to a unique p-value.

The analysis is only performed in blocks in which both wastewater disposal and seismicity occurred during the study period. The p-values from the blocks in each state combine to yield an overall statewide p-value. The aggregated p-values are equal to the probability of the observed association between injection and seismicity (assuming that the null hypothesis is true) in the blocks where the analysis is performed (blocks containing both injection and seismicity). If the p-value is sufficiently low, we can assess statistical confidence that injection is associated with seismicity in these blocks.

As the blocks with both injection and seismicity that we study are a subset of the entire state, the aggregated p-values for the blocks in each state can be interpreted more broadly as assessing confidence in the statement: "seismicity associated with wastewater injection has occurred somewhere in the state during the study period." In both states, there are large number of blocks that contain injection but no seismicity. Obviously, in these blocks, there has not been induced seismicity, and these blocks are not included in the analysis. Therefore, the statewide p-values that we calculate do not quantify the probability that any particular injection well in the state will be associated with seismicity; they express confidence in whether annual injection volumes are associated with seismicity somewhere in the state.

As discussed in Section 1.3, association does not necessarily imply causation because of the presence of confounding variables. This is a significant problem for cross-sectional studies of induced seismicity because seismicity and injection are not located randomly (and so associations could be caused by an underlying causal process). However, our study seeks association across time-series within each block. In this setup, under two assumptions our results may be interpreted as causal. The first assumption is that blocks are comparable units before and when wastewater injection starts. This assumption also underlies our method to finding correlations and is plausible given the decades long study period relative to geologic time spans. The second assumption required is with in each block there is no mechanism that simultaneously alters both human wastewater injection and natural seismicity.

Very recently, operators in Oklahoma have begun shutting in wells in response to seismicity, creating the potential for a causal relationship that would tend to reduce injection rates in years when earthquakes occur (whether or not they are natural or induced). This could have a more complex effect over multiple year periods if injection is subsequently restarted after worries subside. However, to our knowledge, there was only a single injection curtailment in response to seismicity in Oklahoma during the study period (2000 - 2013), a well that was shut-in in September 2013 (Oklahoma Corporation Commission, 2016), and there has never been curtailment of wastewater disposal in response to seismicity in California.

In an ideal study, our baseline would be estimated in each region using seismicity data from before the start of injection. This is not possible as injection predates the start of the seismicity record. Instead, we estimate the baseline seismicity in each block over the entire time of observation, which allows some of effect of injection could be capture by the baseline. This may lead to underestimated effects.

The ideal treatment variable in the analysis would be induced change in Coulomb stress as a function of position, depth, and time everywhere in the subsurface. Increases in Coulomb stress above the highest point previously reached at a location should be expected to correlate with increased seismicity. With detailed knowledge of subsurface structure and properties, it is possible to calculate changes in pressure based on knowledge of injection volume (Shirzaei et al., 2016). However, for statistical studies across large regions, data availability is not sufficient for this to be possible. We have chosen to use annual injection volume as a reasonable first-order proxy for the treatment variable. In future work, more sophisticated treatment variables could be used based on physical models relating injection volume to changes in Coulomb stress.

Equation 2 assumes that injection only affects seismicity in the year that it occurs. It is possible that injection increases induced seismicity hazard over a longer time scale. This hypothesis could be tested in future work using the methodology outlined in this paper. It would not be possible to account for the potential that seismicity observations could be affected by injection prior to the beginning of the dataset.

We test only for relationships between wastewater injection volumes and seismicity. However, reservoir depletion and injection for improved oil recovery could also cause seismicity. Water coproduced with oil and gas is usually reinjected in disposal wells, and so wastewater disposal tends to be associated with other oil and gas activities. Therefore, our study does not causally identify seismicity induced only by wastewater disposal, but rather finds seismicity induced by wastewater disposal and all other associated human activities.

To calculate the statewide p-values, we take the product of the p-values in each block. If the null hypothesis is true, the p-values will be uniformly distributed between 0 and 1. Statewide results can be considered significant if the overall distribution of p-values deviates sufficiently from a uniform distribution. The negative of the natural logarithm of the product of $N$ random variables uniformly distributed between 0 and 1 is distributed according to the gamma distribution with shape factor equal to $N$ and scale factor equal to 1 (page 405 from Devroye, 1986). This property allows for computing overall p-values considering all the blocks in each state.

For blocks with $D$ equal to zero (p-value between 0.5 and 1.0), the p-value is set to $(1.0 + n_0/m)/2$, where $n_0$ is the number of resampled datasets for which $D$ is not equal to zero and $m$ is the total number of resamples. This results in a unique calculation of statewide p-value that is close to the overall expected value of the p-value.

### 2.3. Calculating the maximum likelihood estimates.

In order to estimate the likelihood value in block $i$ for a particular value of $\beta_i$ and $\mu_i$, we perform the integral:

$$P(y_i|x_i, \mu_i, \beta_i, a, \sigma_{II}) = \int P(y_i|x_i, \mu_i, \sigma_i, \beta_i, a, \sigma_{II}) P(\sigma_i) d\sigma_i. \tag{4}$$

Evaluating this integral requires estimation of $P(y_i|x_i, \mu_i, \sigma_i, \beta_i, a, \sigma_{II})$. The prior for $P(\sigma_i)$ is uniformly distributed between 0.01 and 10, which covers a very broad range.

For fixed $\mu_i, \beta_i,$ and $\sigma_i$, the likelihood in each year, $P(y_{ij}|x_{ij}, y_{i,j-1}, \mu_i, \sigma_i, \beta_i, a, \sigma_{II})$, can be calculated by numerically integrating over the Normal distributions in Equation 2. The total likelihood is the product of the observations from each year. This assumes that the $a$ and $\sigma_{II}$ term has accounted for temporal correlation.

$$P(y_i|x_i, \mu_i, \sigma_i, \beta_i, a, \sigma_{II}) = \prod_{j=1}^{N_{years}} \int \int poisspdf(y_{ij}, (\mu_i + \beta_i x_{ij})e^{X_1} + ay_{i,j-1}e^{X_2}) normpdf(X_1, 0, \sigma_i) normpdf(X_2, 0, \sigma_{II}) dX_1 dX_2, \tag{5}$$

where $normpdf(X, 0, \sigma)$ evaluates the probability density function of a Normal distribution at value $X$, mean of 0, and standard deviation $\sigma$, and $poisspdf(y, \mu)$ evaluates the probability mass function of the Poisson distribution at value $y$ and rate parameter $\mu$.

The integrals in Equations 4 and 5 are performed with the code Cubature (Johnson, 2013), which uses the adaptive cubuture method described by Genz and Malik (1980).

The maximum likelihood values of $\beta_i$ and $\mu_i$ are estimated with a brute-force grid search, an approach that is computationally feasible because of the low dimensionality of the problem. The grid search is performed on the logarithm of the parameters. The analysis starts with initial bounds on the parameters that are chosen to cover the full range of feasible values. The search space is divided into a uniform 20×20 grid, and the likelihood is calculated at each point on the grid. The point with highest likelihood is selected, and then the grid search is repeated, refining the grid bounds after each step to search in the neighborhood of the best result from the previous step. The algorithm stops when the change in log-likelihood from one step to another is less than $10^{-4}$. This approach reduces the risk of converging to local minima because it initially samples uniformly across the entire search domain.

**2.4. Estimating temporal correlation parameters.** The optimization of $a$ and $\sigma_{II}$ is performed in the 531 blocks in California that had seismicity but no wastewater disposal. The likelihood function in each block is evaluated as:

$$P(y_i|x_i, a, \sigma_{II}) = \int \int P(y_i|x_i, \mu_i, \sigma_i, a, \sigma_{II}) P(\sigma_i) P(\mu_i) d\sigma_i d\mu_i. \qquad (6)$$

The prior distribution for $\sigma_i$ is assumed uniform between 0.01 and 5.0, and the prior distribution for $\mu_i$ is assumed log-uniform between $10^{-8}$ and 10. The likelihood function across the entire state is the product of the likelihood function from each block. $P(y_i|x_i, \mu_i, \sigma_i, a, \sigma_{II})$ is evaluated from Equation 5. $\beta_i$ is neglected in this optimization because only blocks are used that do not contain any wastewater injection.

The integration in Equation 6 is performed by using a Gibbs sampler (Gelman et al., 2004) to draw values from the posterior distribution $P(\mu_i, \sigma_i|y_i, x_i, a, \sigma_{II})$ and then taking the arithmetic average of the likelihood values. The maximum likelihood estimators for $a$ and $\sigma_{II}$ are calculated using the iterative, stochastic gradient Adagrad algorithm (Duchi et al., 2011). Results from the Gibbs sampler are used in the posterior predictive check described in Section 3.1.

## 3. Results and discussion.

**3.1. Posterior predictive check.** We use posterior predictive checking, a resampling based mechanism (Gelman et al. 2004), to test the ability of Equation 2 to represent the data. Posterior predictive checks are performed by replicating data from the model using the posterior predictive distribution. The replicated data are compared to the observed data to test whether they are similar.

We perform a posterior predictive check with the 531 California blocks with seismicity but no injection. A Gibbs sampler is used to draw 150 samples of $\mu_i$ and $\beta_i$ from the posterior in each block, $P(\mu_i, \sigma_i|y_i, x_i, a, \sigma_{II})$, and then ten forward simulations of the data, $y_i$, are performed for each combination of $\mu_i$ and $\beta_i$ drawn from the posterior. The results from all blocks are aggregated to calculate the frequency distribution of $y_{ij}$ values in the simulated data, which are compared to the real data. Figure 1 shows the $y_{ij}$ frequency plot.

Figure 1 shows that the aggregated simulation data closely matches the observations. At high values of $y_{ij}$, the frequency becomes very low and the sample size in the data becomes insufficiently large to accurately calculate the frequency. This causes the curve to become discontinuous and level out at a minimum value equal to one divided by the total number of samples.

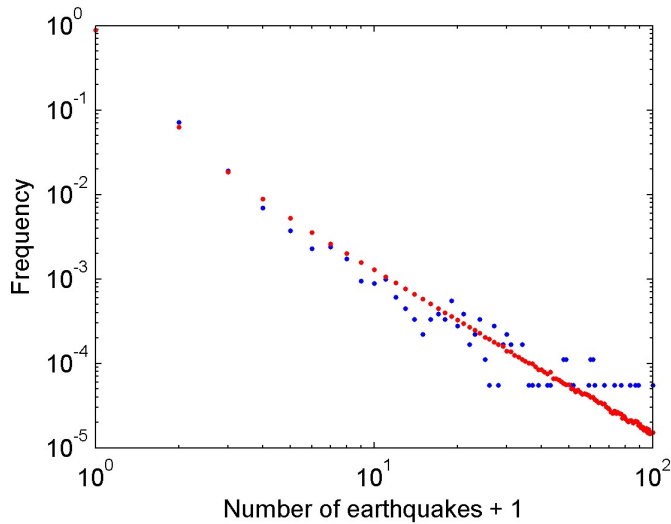

Fig. 1. *Aggregate frequency of $y_{ij}$ in the 531 California blocks with seismicity but no wastewater disposal, both actual data (blue) and simulations from the posterior (red). The x-axis shows one plus the total number of earthquakes, so that zero earthquakes is plotted on the x-axis at $10^0$.*

**3.2. California.** Figure 2 shows maps of cumulative wastewater injection volume, cumulative number of earthquakes, the p-value in blocks with both injection and seismicity, and the maximum likelihood $\beta_i$ in blocks with p-value below 0.3. The complete results are given in Table 1. Figure 3 shows the cumulative distribution function of p-values for blocks in the state. There are not any blocks for which the calculated $D$ was greater than all of $D$ values from the resampled datasets. The analysis in this section

The overall statewide p-value is 0.087. This p-value is higher than the traditional 0.05 threshold for statistical significance, but does indicate that seismicity induced by wastewater disposal (or other spatially associated activities such as reservoir depletion) has probably occurred somewhere in the state. Three blocks (6% of the blocks with both injection and seismicity) have p-value below 0.05, which is consistent with what would be expected from chance. However, 16 blocks (31% of blocks with both injection and seismicity) have p-value below 0.2, more than would be expected from chance.

The areas of greatest seismicity do not correlate with the regions of greatest wastewater disposal. Disposal is concentrated in the Central Valley and in some regions near the coast of southern California. Blocks with low p-value appear to be randomly distributed across the state, except for a cluster of low p-value blocks in the vicinity of Coalinga (36.2°N, 120.4°W). Maximum likelihood values for $\beta_i$ are relatively high in some blocks with high p-value. These values are not meaningful because the statistical confidence is so low.

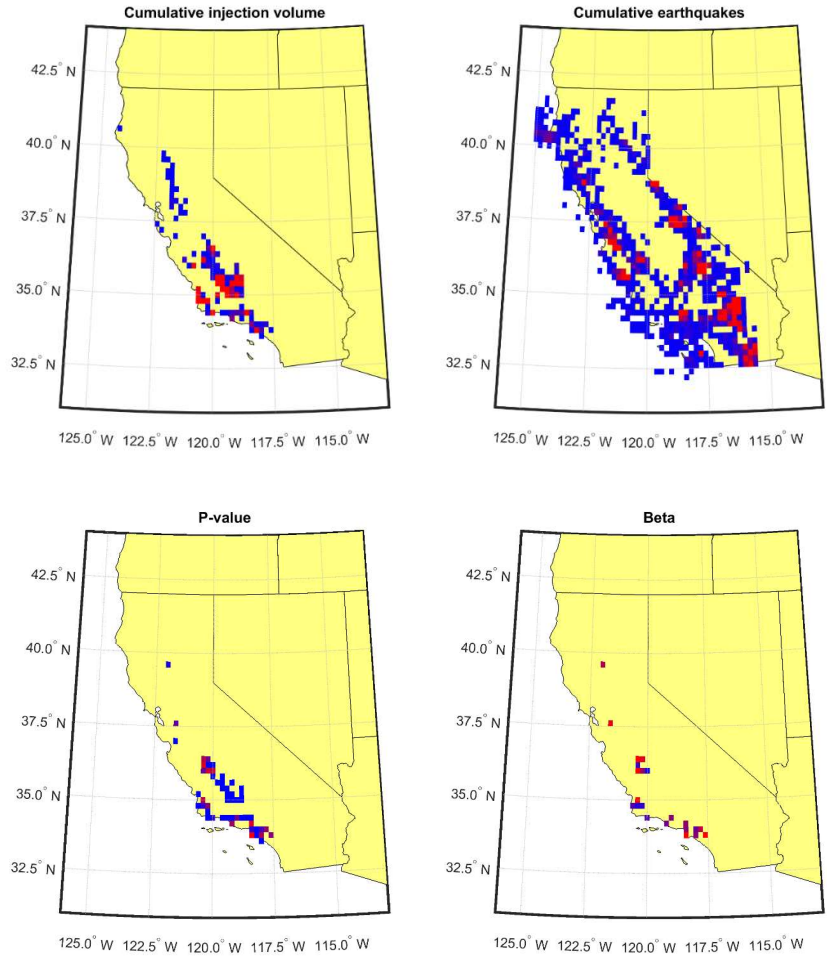

Fig. 2. *First panel: cumulative number of earthquakes shallower than 7 km (bright red representing greater than 50); second panel: cumulative water injected (bright red representing greater than $10^8$ bbl); third panel: p-value for test model including induced seismicity (bright red representing near zero; blue representing 0.3 or higher); fourth panel: maximum likelihood estimate for $\beta_i$ (color scaled logarithmically between $10^{-9}$ and $10^{-6}$; only blocks with p-value less than 0.3 are shown).*

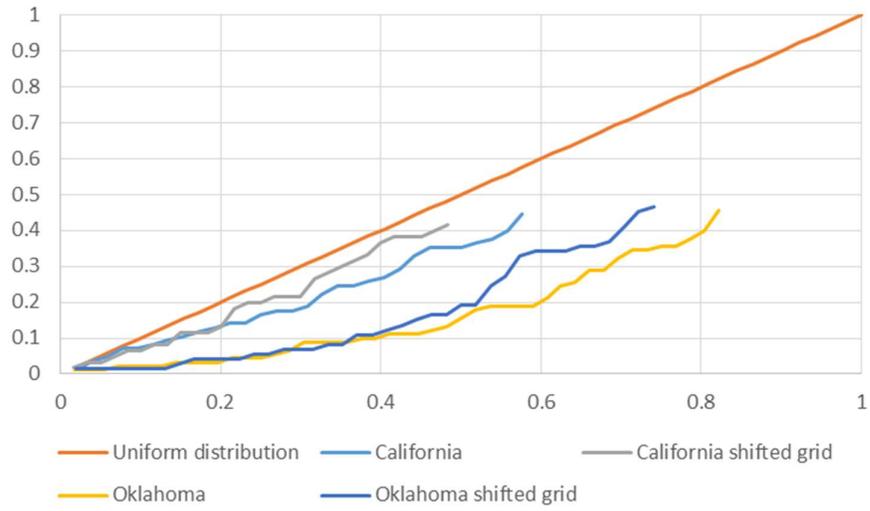

Fig. 3. *Cumulative distribution function of p-values in Oklahoma and California using the original grid and a shifted grid (discussed in Section 3.5). P-values between 0.5 and 1.0 (corresponding to D equal to zero) are not shown.*

Table 1: *Summary of data and results in California.*

| Latitude of block center | Longitude of block center | P-value | Cumulative earthquakes | Cumulative wastewater disposal (bbl) | Optimal $\mu_i$ with $\beta_i$ set to zero | Optimal $\mu_i$ with the optimal $\beta_i$ | Optimal $\beta_i$ |
|---|---|---|---|---|---|---|---|
| 33.8 | 118.4 | 0.012 | 4 | 1.97E+06 | 7.08E-02 | 8.74E-10 | 1.32E-06 |
| 36.2 | 120.4 | 0.035 | 43 | 1.05E+08 | 3.48E-03 | 8.74E-10 | 5.53E-09 |
| 34.2 | 118.4 | 0.047 | 25 | 3.05E+05 | 5.78E-07 | 8.47E-10 | 5.05E-08 |
| 36 | 120.2 | 0.071 | 2 | 1.95E+07 | 3.10E-02 | 8.74E-10 | 5.63E-08 |
| 36 | 120 | 0.071 | 15 | 1.16E+07 | 4.13E-05 | 8.47E-10 | 1.49E-09 |
| 35 | 120.4 | 0.082 | 1 | 2.64E+05 | 8.39E-03 | 8.47E-10 | 1.27E-06 |
| 36.4 | 120.4 | 0.094 | 26 | 3.22E+05 | 2.15E-02 | 9.78E-03 | 1.67E-06 |
| 33.8 | 118 | 0.106 | 2 | 3.10E+07 | 3.10E-02 | 8.47E-10 | 3.65E-08 |
| 34.2 | 119.2 | 0.118 | 2 | 4.91E+07 | 3.10E-02 | 9.48E-10 | 2.28E-08 |
| 34 | 117.8 | 0.129 | 6 | 1.43E+07 | 4.74E-02 | 8.74E-10 | 1.68E-07 |
| 33.8 | 117.6 | 0.141 | 4 | 4.00E+03 | 3.89E-02 | 2.94E-02 | 8.79E-05 |
| 34.8 | 120.2 | 0.141 | 5 | 3.17E+08 | 1.11E-02 | 1.18E-09 | 3.60E-09 |
| 34.4 | 119 | 0.165 | 2 | 2.81E+07 | 3.10E-02 | 9.48E-10 | 4.00E-08 |
| 36 | 120.4 | 0.176 | 7 | 1.16E+05 | 8.39E-02 | 4.41E-02 | 9.09E-06 |
| 36.4 | 120.2 | 0.176 | 5 | 1.22E+06 | 1.09E-02 | 2.87E-03 | 3.77E-07 |
| 37.6 | 121.6 | 0.188 | 15 | 1.13E+06 | 1.24E-02 | 1.18E-09 | 5.62E-07 |
| 34 | 118 | 0.224 | 1 | 7.31E+06 | 8.39E-03 | 8.74E-10 | 1.89E-08 |
| 34 | 118.4 | 0.247 | 9 | 5.17E+07 | 1.94E-01 | 1.40E-01 | 3.89E-08 |
| 34.8 | 120.4 | 0.247 | 5 | 1.06E+09 | 8.22E-02 | 1.69E-03 | 2.73E-09 |
| 34.4 | 120 | 0.259 | 2 | 1.21E+08 | 3.10E-02 | 8.47E-10 | 8.62E-09 |
| 39.6 | 122 | 0.271 | 1 | 2.46E+06 | 8.39E-03 | 8.74E-10 | 1.14E-07 |
| 34.8 | 120.6 | 0.294 | 3 | 2.30E+08 | 5.68E-02 | 4.19E-03 | 7.87E-09 |
| 34.4 | 119.4 | 0.329 | 2 | 3.39E+06 | 3.10E-02 | 2.23E-02 | 8.70E-08 |
| 33.6 | 118 | 0.353 | 3 | 6.78E+05 | 5.68E-02 | 4.41E-02 | 6.47E-07 |
| 34.4 | 118.4 | 0.353 | 110 | 1.52E+07 | 7.04E-02 | 6.06E-02 | 2.64E-08 |
| 35.2 | 119.4 | 0.353 | 2 | 1.15E+09 | 3.10E-02 | 3.11E-02 | 2.09E-12 |
| 37 | 121.6 | 0.365 | 32 | 4.86E+05 | 4.38E-01 | 4.08E-01 | 2.29E-06 |
| 36.2 | 120 | 0.376 | 15 | 1.11E+05 | 2.44E-02 | 1.63E-02 | 2.20E-06 |
| 35 | 119 | 0.400 | 3 | 1.24E+08 | 5.68E-02 | 4.54E-02 | 3.21E-09 |
| 35.4 | 119.6 | 0.447 | 1 | 2.87E+08 | 8.39E-03 | 1.96E-08 | 9.47E-10 |
| 35.2 | 118.8 | 0.659 | 1 | 4.70E+06 | 8.39E-03 | 8.65E-03 | 3.33E-18 |
| 34.4 | 118.8 | 0.665 | 4 | 3.07E+07 | 8.43E-07 | 8.04E-07 | 4.56E-18 |
| 35.6 | 119.2 | 0.671 | 1 | 1.82E+07 | 8.39E-03 | 8.65E-03 | 3.33E-18 |
| 34.6 | 120.2 | 0.676 | 1 | 9.21E+05 | 8.39E-03 | 8.65E-03 | 3.33E-18 |
| 35.6 | 119.8 | 0.676 | 1 | 5.16E+08 | 8.39E-03 | 8.65E-03 | 3.33E-18 |
| 35.4 | 118.8 | 0.682 | 3 | 1.02E+09 | 4.04E-02 | 3.81E-02 | 4.56E-18 |
| 35.6 | 119.6 | 0.682 | 1 | 4.69E+08 | 8.39E-03 | 8.65E-03 | 3.33E-18 |
| 34.4 | 119.6 | 0.688 | 3 | 1.51E+04 | 5.68E-02 | 5.62E-02 | 4.56E-18 |
| 35.4 | 119.4 | 0.694 | 4 | 2.93E+06 | 8.45E-02 | 8.29E-02 | 4.56E-18 |
| 35 | 118.8 | 0.700 | 6 | 1.78E+07 | 6.80E-07 | 6.31E-07 | 4.56E-18 |
| 35 | 119.4 | 0.706 | 1 | 4.54E+08 | 8.39E-03 | 8.65E-03 | 3.33E-18 |
| 35.2 | 119.2 | 0.706 | 1 | 6.27E+07 | 8.39E-03 | 8.65E-03 | 3.33E-18 |
| 36.2 | 120.2 | 0.718 | 49 | 6.66E+06 | 1.97E-03 | 1.97E-03 | 3.33E-18 |
| 34.4 | 118.6 | 0.724 | 104 | 4.53E+08 | 5.75E-02 | 5.62E-02 | 3.67E-13 |
| 35 | 119.2 | 0.724 | 7 | 6.87E+07 | 2.40E-02 | 2.39E-02 | 2.96E-18 |
| 34.4 | 119.2 | 0.729 | 3 | 2.24E+07 | 1.08E-06 | 1.10E-06 | 3.33E-18 |
| 35.2 | 120.6 | 0.729 | 2 | 9.28E+07 | 3.10E-02 | 3.11E-02 | 4.18E-11 |
| 35.8 | 120 | 0.735 | 2 | 1.35E+05 | 1.77E-06 | 1.78E-06 | 4.56E-18 |
| 34.4 | 120.2 | 0.753 | 1 | 1.05E+07 | 8.39E-03 | 8.65E-03 | 3.33E-18 |
| 35.8 | 119.6 | 0.753 | 4 | 1.91E+04 | 7.41E-03 | 7.08E-03 | 4.56E-18 |
| 33.8 | 118.2 | 0.759 | 7 | 7.83E+08 | 8.21E-03 | 8.19E-03 | 3.33E-18 |
| 34 | 118.2 | 0.765 | 4 | 4.39E+06 | 5.22E-04 | 5.19E-04 | 3.33E-18 |

One block, centered at (33.8°N, 118.4°W) in Rancho Palos Verdes, has a particularly low p-value, 0.012. There were only four earthquakes in this block over the entire 34 year study period, all with magnitude below 4.0 and occurring between 1986 and 1991 (Figure 4). The injection wells were located in Hermosa Beach and the earthquakes were a few miles away off the coast. The analysis finds a low p-value because the only wastewater disposal in this block occurred from 1986 to 1992, a period of time coinciding with the observed seismicity.

The correlation suggests that there is an association between injection and seismicity in the block. However, to be conclusive, the association needs to be confirmed with a detailed site-specific study (such as the study performed by Hornbach et al., 2015).

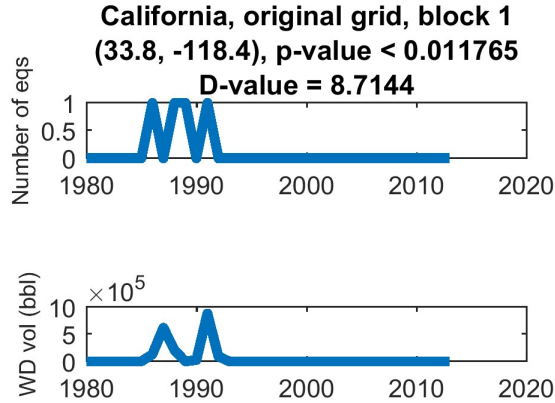

Fig. 4. *Total earthquakes and wastewater disposal volume per year in the block centered at (33.8°N, 118.4°W).*

The block with the second-lowest p-value, 0.035, is centered at (36.2°N, 120.4°W), near Coalinga, CA (Figure 2). The seismicity in this block was dominated by the aftershocks of the magnitude 6.2 Coalinga earthquake in 1983. The mainshock had a depth of 10 km, and our dataset includes only earthquake with depth 7 km or less. However, many of the aftershocks were shallower than 7 km, and so our dataset includes 38 earthquakes in the block in 1983. The algorithm identifies a potential relationship because wastewater injection volumes declined significantly after 1994, and there have not been any earthquakes in the block since 1993. Figure 2 shows that the surrounding blocks also have relatively low p-values. The surrounding blocks have similar observations: that there was significantly more seismicity and wastewater injection prior to the mid-1990s. The seismicity near Coalinga, CA in the 1980s has previously been identified as potentially induced by fluid extraction (which is associated with wastewater injection). Detailed mechanistic investigations have come to differing conclusions about whether induced seismicity is a plausible hypothesis (Segall, 1985; McGarr, 1991).

**3.3. Oklahoma.** Figure 5 shows that there has been a large increase in seismicity in Oklahoma in recent years. Injection volumes increased in the state throughout the 2000s, but the increase in seismicity did not begin until 2008. Complete injection volume data is not available prior to the late 1990s.

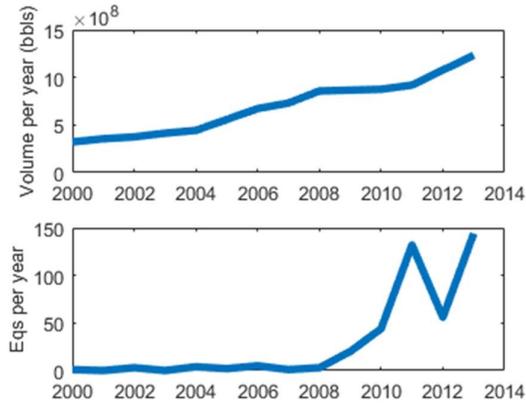

Fig. 5. *Cumulative volume of wastewater disposal injection and number of earthquakes above magnitude 2.9 per year in Oklahoma.*

Figure 6 shows cumulative water injection, cumulative number of earthquakes, and p-value for induced seismicity across Oklahoma. The raw data is given in Table 2. The cumulative distribution function of the block p-values is shown in Figure 3. Wastewater injection is broadly distributed across the state. The largest injection volumes have occurred in a crescent-shaped region around Oklahoma City (35.5°N, 97.5°W). Seismicity is primarily located along a north-south strip in the central part of the state, with the most significant seismicity occurring to the east of Oklahoma City. Blocks with low p-value are predominantly located to the east of Oklahoma City and in a region near the Kansas border.

Some blocks in the state have low maximum likelihood estimates for $\beta_i$ (in the range of $10^{-9}$) even though they have low p-values. This occurs because these blocks have very high injection volumes and there is temporal association of injection and seismicity, but the number of events relative to the injection volumes is relatively low.

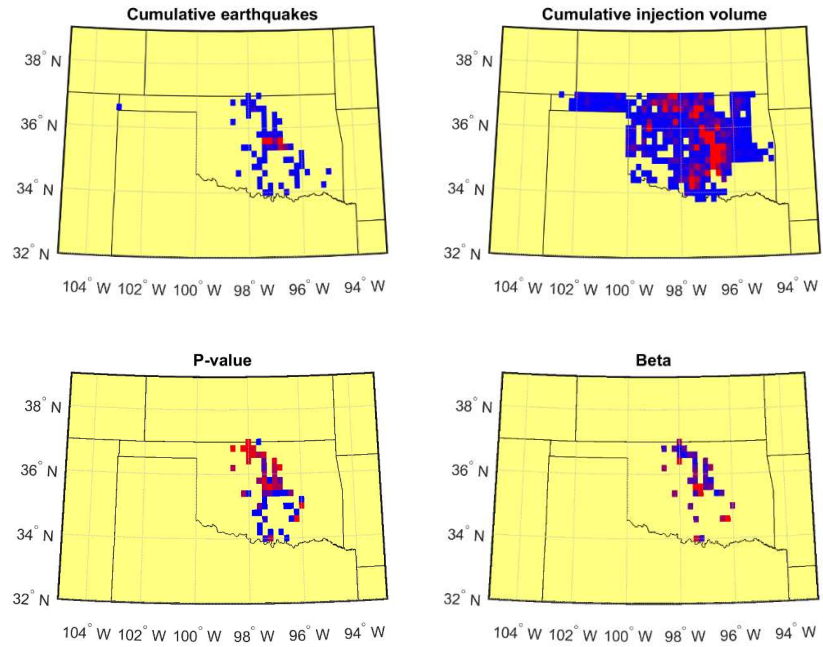

Fig. 6. *First panel: cumulative number of earthquakes (bright red representing greater than 50); second panel: cumulative water injected (bright red representing greater than $10^8$ bbl); third panel: p-value for test model including induced seismicity (bright red representing zero; blue representing 0.3 or higher); fourth panel: maximum likelihood estimate for $\beta_i$ (color scaled logarithmically between $10^{-9}$ and $10^{-6}$; only blocks with p-value less than 0.3 are shown).*

Table 2. *Summary of results and data in Oklahoma.*

| Latitude of block center | Longitude of block center | P-value | Cumulative earthquakes | Cumulative wastewater disposal (bbl) | Optimal $\mu_i$ with $\beta_i$ set to zero | Optimal $\mu_i$ with optimal $\beta_i$ | Optimal $\beta_i$ |
|---|---|---|---|---|---|---|---|
| 34.6 | 96.2 | 0.011 | 2 | 1.45E+06 | 1.85E-04 | 8.47E-10 | 8.94E-07 |
| 35.6 | 97.2 | 0.011 | 66 | 3.40E+07 | 4.09E+00 | 2.46E-09 | 1.10E-06 |
| 36.8 | 98.2 | 0.011 | 1 | 4.85E+07 | 1.83E-02 | 8.74E-10 | 1.30E-08 |
| 36.2 | 96.8 | 0.022 | 3 | 3.08E+07 | 3.89E-02 | 9.48E-10 | 6.10E-08 |
| 36.6 | 98 | 0.022 | 1 | 5.80E+07 | 1.83E-02 | 1.18E-09 | 6.31E-09 |
| 36.8 | 98.6 | 0.022 | 1 | 7.72E+07 | 1.83E-02 | 1.18E-09 | 6.31E-09 |
| 37 | 98 | 0.022 | 1 | 7.04E+06 | 1.83E-02 | 1.18E-09 | 7.18E-08 |
| 35 | 96 | 0.033 | 1 | 1.01E+07 | 1.83E-02 | 8.74E-10 | 3.12E-08 |
| 35.8 | 97.4 | 0.033 | 4 | 2.07E+06 | 7.76E-05 | 9.48E-10 | 7.79E-09 |
| 36.6 | 97.8 | 0.033 | 10 | 2.20E+07 | 4.36E-05 | 1.18E-09 | 5.54E-10 |
| 36.8 | 98 | 0.033 | 5 | 2.46E+07 | 6.57E-05 | 9.48E-10 | 1.24E-09 |
| 36.4 | 97 | 0.044 | 2 | 3.28E+07 | 1.85E-04 | 9.48E-10 | 2.28E-08 |
| 36.6 | 97.4 | 0.044 | 2 | 1.34E+08 | 4.30E-02 | 8.47E-10 | 8.53E-09 |
| 36.8 | 97.8 | 0.044 | 2 | 4.80E+07 | 1.85E-04 | 1.18E-09 | 1.61E-08 |
| 36.4 | 98 | 0.056 | 1 | 7.75E+05 | 1.83E-02 | 8.47E-10 | 4.39E-07 |
| 35.6 | 97.4 | 0.067 | 45 | 2.25E+06 | 1.67E-01 | 9.48E-10 | 1.77E-06 |
| 35.6 | 97 | 0.089 | 22 | 1.07E+08 | 7.38E-03 | 9.48E-10 | 8.96E-09 |
| 35.8 | 97 | 0.089 | 5 | 3.65E+08 | 1.27E-01 | 9.48E-10 | 7.79E-09 |
| 36.2 | 97 | 0.089 | 3 | 1.66E+07 | 1.02E-04 | 1.18E-09 | 2.92E-09 |
| 36.2 | 98.6 | 0.089 | 1 | 4.44E+06 | 1.83E-02 | 8.47E-10 | 5.03E-08 |
| 34 | 97.2 | 0.100 | 3 | 1.52E+06 | 1.02E-04 | 9.48E-10 | 5.53E-09 |
| 36.6 | 97.6 | 0.100 | 2 | 3.23E+07 | 1.85E-04 | 9.48E-10 | 1.37E-08 |
| 35.4 | 98.2 | 0.111 | 1 | 6.75E+06 | 1.83E-02 | 1.18E-09 | 4.94E-08 |
| 35.6 | 96.8 | 0.111 | 90 | 1.34E+08 | 1.01E-01 | 9.48E-10 | 2.28E-08 |
| 36 | 97 | 0.111 | 1 | 3.33E+07 | 1.83E-02 | 1.18E-09 | 1.11E-08 |
| 36 | 97.4 | 0.122 | 1 | 6.25E+06 | 1.83E-02 | 9.48E-10 | 6.10E-08 |
| 35.8 | 97.6 | 0.133 | 1 | 2.99E+07 | 1.83E-02 | 1.18E-09 | 1.11E-08 |
| 35.4 | 96.4 | 0.156 | 4 | 1.02E+08 | 2.18E-02 | 8.74E-10 | 9.99E-09 |
| 35 | 97.6 | 0.178 | 1 | 1.83E+07 | 1.83E-02 | 1.18E-09 | 1.61E-08 |
| 35.4 | 97.2 | 0.189 | 11 | 5.45E+06 | 3.96E-01 | 1.58E-01 | 5.62E-07 |
| 35.4 | 97.4 | 0.189 | 5 | 4.09E+08 | 1.07E-01 | 8.74E-10 | 4.81E-09 |
| 35.8 | 96.8 | 0.189 | 1 | 1.76E+08 | 1.83E-02 | 9.48E-10 | 1.74E-09 |
| 36.4 | 97.4 | 0.189 | 1 | 1.60E+08 | 1.83E-02 | 8.74E-10 | 2.11E-09 |
| 36.2 | 97.4 | 0.211 | 1 | 3.27E+07 | 1.83E-02 | 8.74E-10 | 9.70E-09 |
| 34 | 97.4 | 0.244 | 2 | 5.53E+06 | 7.83E-02 | 1.18E-09 | 2.21E-07 |
| 34.6 | 97.4 | 0.256 | 1 | 1.65E+07 | 1.83E-02 | 1.18E-09 | 1.61E-08 |
| 34.6 | 96.4 | 0.289 | 2 | 1.36E+07 | 7.83E-02 | 9.48E-10 | 8.86E-08 |
| 35.4 | 96.6 | 0.289 | 23 | 3.72E+08 | 1.72E-03 | 1.18E-09 | 3.16E-10 |
| 35 | 96.4 | 0.322 | 1 | 8.07E+07 | 1.83E-02 | 1.18E-09 | 3.60E-09 |
| 35 | 96.6 | 0.344 | 1 | 1.24E+08 | 1.83E-02 | 1.18E-09 | 2.48E-09 |
| 35.4 | 96.8 | 0.344 | 38 | 1.62E+08 | 3.98E-05 | 9.48E-10 | 4.05E-12 |
| 34.8 | 97.8 | 0.356 | 2 | 1.95E+07 | 4.30E-02 | 9.48E-10 | 3.48E-08 |
| 35.2 | 97.4 | 0.356 | 2 | 8.68E+06 | 1.85E-04 | 9.48E-10 | 3.61E-10 |
| 35.2 | 96 | 0.378 | 1 | 5.40E+06 | 1.83E-02 | 1.18E-09 | 4.94E-08 |
| 35 | 97.4 | 0.400 | 1 | 1.13E+07 | 1.83E-02 | 8.74E-10 | 2.47E-08 |
| 34.2 | 97.6 | 0.456 | 2 | 1.39E+08 | 1.85E-04 | 9.48E-10 | 1.96E-11 |
| 34.8 | 96.2 | 0.694 | 1 | 2.68E+07 | 1.83E-02 | 1.82E-02 | 2.96E-18 |
| 35.4 | 97 | 0.700 | 6 | 2.99E+06 | 7.46E-02 | 7.42E-02 | 2.96E-18 |
| 35.4 | 98 | 0.700 | 1 | 5.10E+06 | 1.83E-02 | 1.82E-02 | 2.96E-18 |
| 34.4 | 97.6 | 0.711 | 1 | 1.87E+08 | 1.83E-02 | 1.82E-02 | 2.96E-18 |
| 37 | 97.6 | 0.711 | 1 | 3.82E+06 | 1.83E-02 | 1.82E-02 | 2.96E-18 |
| 34.2 | 96.8 | 0.728 | 3 | 4.69E+05 | 5.89E-02 | 5.62E-02 | 4.56E-18 |
| 35.8 | 97.2 | 0.728 | 5 | 2.73E+08 | 8.91E-02 | 8.90E-02 | 2.96E-18 |
| 34 | 96.6 | 0.733 | 5 | 1.73E+06 | 6.57E-05 | 6.58E-05 | 3.33E-18 |
| 34.8 | 97.6 | 0.7611111 | 2 | 36490780 | 0.0001845 | 0.0001813 | 3.33E-18 |
| 34.2 | 97.2 | 0.7666667 | 1 | 2726859 | 0.0182894 | 0.0182158 | 2.96E-18 |

The overall p-value for induced seismicity in Oklahoma is 3.7e-09, indicating very high statistical confidence that there is a relationship between wastewater disposal and seismicity in the state. This result is consistent with other studies, which found strong evidence that the induced seismicity in Oklahoma has been induced by wastewater disposal (Keranen et al., 2014; Walsh and Zoback, 2015).

In three blocks, there were not any resampled datasets with $D$ greater than observed in the data, and so the reported block p-values are only estimates of the upper bound. If these p-values were known more accurately, the estimated state p-value would be even lower.

The block with highest $D$ value is centered at (35.6°N, 97.2°W), the north-eastern part of Oklahoma City. This area was identified as having induced seismicity in a detailed study by Keranen et al. (2014). The data for this block is shown in Figure 7.

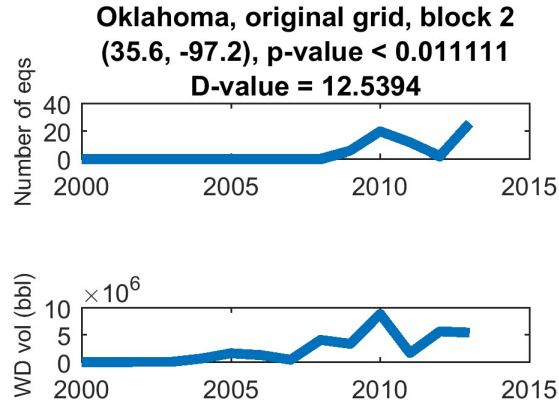

Fig. 7. *Total earthquakes and wastewater disposal volume per year in the block centered at (35.6°N, 97.2°W).*

**3.4. Results with a shifted grid.** To test the model sensitivity to the grid, the analysis is rerun in both states with the grid shifted 0.1° latitude and longitude to the southeast, so that the centers of the blocks in the new grid are located at the corners of the original grid. The results, which are provided in the Supporting Information, are similar to the results from the original analysis.

With the shifted grid, the California p-value is 0.28, instead of 0.087. There is one block in which the calculated $D$ value is greater than the $D$ value from all resampled datasets (60 in this case). In this block, the p-value estimate (0.0167) is an upper bound. If it were known with more accuracy, the statewide p-value would be lower. However, the difference in $D$ values between the data and the smallest resampled $D$ is small, and so it would likely not make a large difference.

25% of the blocks have p-value below 0.2, slightly greater than would be expected from chance. Two of the three lowest p-value blocks in the shifted California analysis are located near Coalinga and Hermosa Beach, at 0.1° shifts from the blocks with lowest p-value in the original analysis. In the shifted grid, there is a new block with an especially low p-value, less than 0.0167, located at (36.1°N, 120.1°W), east of Coalinga (Figure 8).

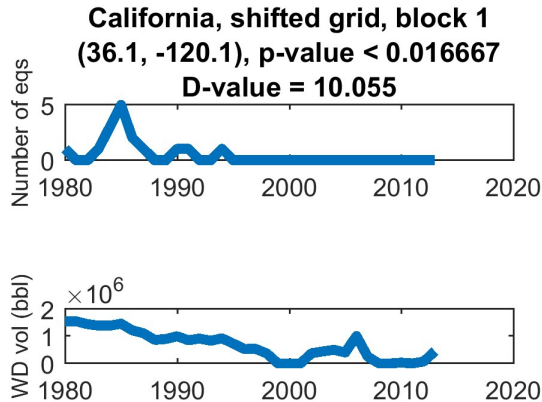

Fig. 8. *Total earthquakes and wastewater disposal volume per year in the block (using the shifted grid) centered at (36.1°N, 120.1°W).*

In Oklahoma, the statewide p-value is 2.5e-07 with the shifted grid, indicating very strong statistical significance (consistent with the results from the original grid). Seven blocks had $D$ value greater than the largest $D$ observed in the resampled datasets (73 in this case), which means that their calculated p-values are only upper bounds. If these p-values were known with more accuracy, the statewide p-value would be even lower. The block with highest $D$ value is at (35.5°N, 97.1°W), shifted 0.1° from the block with the lowest p-value with the original grid. In the shifted data, the lowest p-values are clustered in the northern part of the state and east of Oklahoma City, consistent with the result from the original analysis.

**3.5. Limitations of the method.** Our analysis has some limitations for identifying induced seismicity in individual blocks.

Equation 2 considers only temporal correlation in consecutive years. It is possible that temporal correlations in natural seismicity exist over periods longer than one year. If natural seismicity shows temporal correlations over longer time periods, it would weaken the statistical significance of our results because multi-year sequences of elevated seismicity, which tend to be located in low p-value blocks (Figures 4, 7, and 8), would be more likely to occur by chance.

A second limitation is that the use of gridblocks causes a discretization effect. Associations between injection and seismicity may be missed if they occur across gridblock boundaries. This issue is mitigated in our study because the gridblocks are relatively coarse (approximately 14×11 mi$^2$). Also, we tested the discretization effect by shifting the grid and repeating the calculations. If our method was applied with a finer spatial discretization, it would be more important to account for this discretization effect.

Because of these issues, there is some imprecision in the calculations of p-values. The analysis can identify regions where an association is possible or probable, but the suspected relationships need to be confirmed with detailed site-specific study, such as described by Frohlich (2012), Hornbach et al. (2015), and Segall (1985). These investigators looked in detail at the location and depth of human activities and the location and timing of the earthquake hypocenters.

Our analysis is performed at a relatively coarse spatial and temporal scale and specifically investigates relationships between injection volume and seismicity. Our analysis could miss minor instances of induced seismicity that would be apparent from more detailed analysis. However, any attempt at identifying induced seismicity from more detailed data must carefully account for the possibility of coincidental association due to the large number of hypotheses tested and must model the dependencies between the more detailed units. Our general approach of empirically deriving p-values – using an ensemble of resampled datasets created by permuting and temporally offsetting injection data – can be used with nearly any method of statistically identifying associations between human activity and seismicity that uses a longitudinal study design.

There may be seismicity induced by other activities, such as depletion, exploitation of geothermal energy, and injection for improved oil recovery, that would be not be detected by our analysis unless these activities were associated with wastewater disposal injection.

**3.6. Extensions of the method.** By modifying Equation 1, our method could be extended to test for relationships between induced seismicity and geological parameters such as lithology of injection interval, proximity to faults, or proximity to the granitic basement. The method could also be used to test for relationships between induced seismicity and operational parameters. Equation 1 could be replaced with any statistical model of earthquake behavior and induced seismicity.

Our method could be applied to search for evidence of induced seismicity in any region where sufficient data is available. However, in many states, this type of regional study is hampered by poor data availability. Assembling seismicity and injection well data, and making the data publicly available and easy to access, should be a policy priority.

**4. Conclusions.** We perform a longitudinal analysis on years of wastewater disposal and seismicity data from Oklahoma and California. Our method posits a statistical relationship between injection and seismicity that accounts for temporal correlations using parameters inferred via maximum likelihood estimates on blocks in California with seismicity with no wastewater disposal. We compute the maximum likelihood estimate for a parameter that expresses the relationship between injection and seismicity and use a permutation test to assess the significance of our findings in each block and across each state.

We account for the risk of false-positives due to coincidental association. We also account for nonrandomly located and spatially clustered injection wells and highly temporally variable seismicity observations.

The analysis finds very strong statistical evidence of induced seismicity in Oklahoma associated with wastewater injection, consistent with the results from prior studies. Induced seismicity is most significant in the central and northern part of the state.

The analysis indicates that there has probably been induced seismicity in California associated with wastewater injection, but the results are not conclusive. The analysis identifies an area of potentially induced seismicity near Hermosa Beach, CA where more detailed study would be justified. Also, it identifies earthquakes in the vicinity of Coalinga, CA in the mid-1980s that have previously been identified in the literature as being potentially induced. The analysis identifies other areas in California where induced seismicity is possible and stronger association may be apparent from more detailed analysis.

**Acknowledgements**. Thank you to Matthew Weingarten and Rall Walsh for providing their databases of Oklahoma injection well data. Thank you to Jef Caers for providing useful comments on an earlier version of the manuscript, especially on how to estimate p-values. Thank you to Norm Sleep for suggesting that we test the sensitivity to the grid. The financial support of the Southern California Earthquake Center (Contribution No. 7118), The Cockrell School of Engineering at The University of Texas, the Seibel foundation, and the Porter Ogden Jacobus foundation is gratefully acknowledged. The Southern California Earthquake Center is funded by NSF Cooperative Agreement EAR-1033462 & USGS Cooperative Agreement G12AC20038. The sources of the raw data used for the study are described in the text. Please contact the authors for a copy of the processed data ($x$ and $y$ time series in each block), for the code used to perform the analysis, or for a copy of the Supporting Information.